\documentstyle[aps,twocolumn]{revtex}
\newcommand{\bea}{\begin{eqnarray}}
\newcommand{\beq}{\begin{equation}}
\newcommand{\eea}{\end{eqnarray}}
\newcommand{\eeq}{\end{equation}}
\begin{document}
\title{Hidden symmetries of two-electron
quantum dots in  a magnetic field}
\author{ N. S. Simonovi\'c $^{1}$ and R.G. Nazmitdinov $^{1,2,3}$}
\address{$^{1}$ Max-Planck-Institut f\"ur Physik komplexer
Systeme, D-01187 Dresden, Germany\\
$^{2}$ Departament de F{\'\i}sica, Universitat de les Illes Balears, E-07071
Palma de Mallorca, Spain\\
$^{3}$ Bogoliubov Laboratory of Theoretical Physics, 
Joint Institute for Nuclear Research, 141980 Dubna, Russia}
\maketitle

\baselineskip 20pt minus.1pt

\begin{abstract}
Using a classical and quantum mechanical analysis,
we show that the magnetic field gives rise to dynamical 
symmetries 
of a three-dimensional axially symmetric 
two-electron quantum dot with a parabolic confinement. 
These symmetries manifest themselves as near-degeneracies 
in the quantum spectrum at specific values of 
the magnetic field and are robust at any strength of the
electron-electron interaction.
\end{abstract}

\vspace{0.2in}
PACS number(s): 03.65.Ge, 73.21.La, 05.45.Mt,  75.75.+a
\vspace{0.2in}

A three-dimensional harmonic oscillator 
with frequencies in rational ratios (RHO)
 and a Coulomb system are
benchmarks for the hidden symmetries which account for 
the accidental degeneracies of 
their quantum spectra (see, e.g., \cite{LL}). 
Recent advances in nanotechnology
create a remarkable opportunity to trace their
dynamical interplay in mesoscopic systems like 
quantum dots, which confine a few electrons to a 
space of a few hundred nanometers.
Indeed, competition between a confining potential, approximated
quite well by the harmonic oscillator (HO), and the repulsive electron-electron
interaction produces a rich variety of phenomena, for example, 
in a two-electron quantum dot (QD) under a 
perpendicular magnetic field (see \cite{kou,mak} and references therein).
In fact, a two-electron QD  
becomes a testing--ground for different 
quantum-mechanical approaches \cite{mak}  
and experimental techniques
that could provide highly accurate data for 
this system \cite{Ash96,kou}.   

If the HO and the Coulomb potential 
are combined, most of the symmetries are expected to be broken.
Nevertheless, in particular cases,
the Coulomb (Kepler) system and the  
RHO may have common  symmetries, as
it was already noticed a long time ago \cite{hill}.
The authors of Ref.~\cite{hill} could not find, however,
a physical application for this phenomenon. 
These symmetries were rediscovered
in the analysis 
of laser--cooled ions in a Paul trap \cite{blum} and 
of the hydrogen atom in the generalized van der Waals potential \cite{alh}. 
The major aspect of the present paper is to demonstrate that 
the hidden symmetries could be observed in
a two-electron QD under a tunable perpendicular 
magnetic field if the effective confinement potential is indeed 
the three--dimensional (3D) HO.
To this aim we focus our analysis  
upon the nonlinear classical dynamics of the system. 
At certain conditions the motion becomes integrable
and this indicates the existence of the symmetries
in the quantum spectrum.   

The Hamiltonian of a two-electron QD in a magnetic
field reads
\beq
\label{ham}
H = \sum_{j=1}^2 \bigg[ \frac{1}{2m^*\!}\,
\Big({\bf p}_j - \frac{e}{c} {\mathbf A}_j \Big)^{\! 2}
+ U({\mathbf r}_j) \bigg]
+ \frac{\alpha}{\vert{\mathbf r}_1 \!- {\mathbf r}_2\vert},
\eeq
where $\alpha = e^2/4\pi\varepsilon_0\varepsilon_r$. 
Here, $e$, $m^*$, $\varepsilon_0$ and
$\varepsilon_r$ are the unit charge, effective electron mass, vacuum
and relative dielectric constants of a semiconductor,
respectively. For the perpendicular magnetic field we choose 
the vector potential with
a gauge ${\mathbf A} = \frac{1}{2} {\mathbf B}
\times {\mathbf r} = \frac{1}{2}B(-y, x,0)$. The confining potential is
approximated by a 3D axially-symmetric HO $U({\mathbf
r}) = m^* [\omega_0^2\,(x^2 \!+ y^2) + \omega_z^2z^2]/2$, where
$\hbar\omega_z$ and $\hbar\omega_0$ are the energy scales of confinement in
the $z$-direction and in the $xy$-plane, respectively.
In contrast to a 2D description of the QD
this approximation provides a consistent description \cite{NSR} of
various experimental features \cite{Ash}: 
the energy spectrum for small magnetic field, the value of the magnetic
field for the first singlet-triplet transition, and the ratio of the lateral
to vertical extension of the dot.
In the present analysis we neglect the spin interaction, since
the corresponding energy is small compared to the confinement
and the Coulomb energies.

Introducing the relative and center-of-mass (CM) coordinates 
${\bf r} = {\bf r}_1 - {\bf r}_2$, $ {\bf R} = 
\frac{1}{2}({\bf r}_1 +  {\bf r}_2)$, the Hamiltonian
(\ref{ham}) is separated into the CM and relative-motion terms
$H = H_{\mathrm CM} + H_{\mathrm rel}$ \cite{Din}. 
The solution to the first term is well
known \cite{Fock}. It possesses all symmetries of the HO,
since the Coulomb interaction enters the second term 
for the parabolic confinement \cite{15}. 
In  the following we concentrate
on the dynamics of $H_{\rm rel}$. For our analysis it is convenient to use
cylindrical {\it scaled} coordinates, $\tilde\rho = \rho/l_0$, ${\tilde
p}_{\rho} = p_{\rho}l_0/\hbar$, $\tilde z = z/l_0$, $ {\tilde p}_{z} = p_z
l_0/\hbar$, where $l_{0} = (\hbar/\mu\omega_0)^{1/2}$ is the characteristic
length of the confinement potential with the reduced mass $\mu = m^*/2$. The
strength parameter $\alpha$ of the Coulomb repulsion goes over to $\lambda =
2\alpha/(\hbar \omega_0 l_0)$. 
Although our consideration is general, for the demonstration
we choose the values
$\hbar \omega_0 \approx 2.8$ meV and $\omega_z = 2.5\, \omega_0$
which are close to those obtained in our 3D analysis \cite{NSR} 
of the experiment \cite{Ash}.
For the effective mass $m^*=0.067 m_e$ and the
dielectric constant $\varepsilon=12$, which are typical for GaAs, 
the value $\lambda = 3$.
Hereafter, for the sake of simplicity,
we drop the tilde, i.e. for the scaled variables we use the same symbols as
before scaling. 

In these variables the  
Hamiltonian for the relative motion takes the form
(in units of $\hbar \omega_0$) 
\beq
h \equiv \frac{H_{\rm rel}}{\hbar\omega_0} = \frac{1}{2}\bigg(
p_\rho^2 + \frac{m^2}{\rho^2} + p_z^2 +
{\tilde\omega_{\!\rho}}^2 \rho^2 + 
{\tilde\omega_z}^2 z^2 +
\frac{\lambda}{r}\bigg) - {\tilde\omega_L} m,
\label{relham}
\eeq
where $r = (\rho^2+z^2)^{1/2}$,
$\tilde\omega \equiv \omega/\omega_0$, $m = l_z/\hbar$.
Here, $\omega_L=eB/2m^*\!c$ is the Larmor frequency and 
$\omega_{\rho}=(\omega_{L}^{2}+\omega_{0}^{2})^{1/2}$
is the effective confinement frequency in the $\rho$-coordinate, which
depends through $\omega_{L}$ on the magnetic field. In this way the
magnetic field can be used to control the effective lateral
confinement frequency  of the QD for a fixed value of the
vertical confinement,
i.e. the ratio $\omega_z/\omega_\rho$.

Beside the energy $(\epsilon \equiv h)$, 
the $z$-component of angular momentum $l_z$
is also an integral of motion
due to the axial symmetry of the system.  
Therefore, the magnetic quantum number $m$ is always a 
good quantum number.
Since the Hamiltonian (\ref{relham}) 
is invariant under the reflection of the origin, 
the parity $\pi$ is a good quantum number too.

Although the motion in $\varphi$ is separated 
from the motion in the $(\rho, z)$-plane,
the problem is in general non-integrable,
since the Coulomb term couples the  $\rho$ and $z$-coordinates.
Examination of the Poincar\'e sections by varying 
the parameter $\omega_z/\omega_\rho$ (see Fig.~\ref{fig1}
for examples) in the interval $(1/10,10)$ with a small step indicates
that there are five integrable cases.
The trivial cases
are $\omega_z/\omega_\rho \to 0$ and $\omega_z/\omega_\rho \to \infty$, which
correspond to 1D circular and 2D vertical QDs, respectively. 
The nontrivial cases are 
$\omega_z/\omega_\rho = 1/2,1,2$.  
These results hold for any strength of the Coulomb
interaction and agree with
the results for the Paul trap \cite{blum}.   
Below we discuss the nontrivial cases only.
The typical trajectories in 
cylindrical coordinates are shown in Figs.~\ref{fig2}a,c.

The results obtained with the aid of the 
Poincar\'e surfaces of sections are invariant under the 
coordinate transformation.
On the other hand, the integrability is a necessary condition for the 
existence of a coordinate system in which the motion can be separated.
In turn, the analogous quantum mechanical system would be characterized 
by a complete set of quantum numbers.

At the value $\omega_L^\prime = (\omega_z^2 - \omega_0^2)^{1/2}$ 
the magnetic field gives rise to a spherical symmetry 
$(\omega_z/\omega_\rho=1)$
in the axially symmetric
QD \cite{NSR}. In this case the Hamiltonian (\ref{relham}) 
is separable in (scaled) spherical coordinates (see Eq.~(7) in \cite{NSR}).
The additional integral of motion is 
the square of the total angular momentum ${\bf l}^2$.
The spherical coordinates 
are a particular limit of the spheroidal (elliptic) 
coordinates well suitable for the analysis
of the Coulomb systems (see, e.g., \cite{Kom}).
Therefore, to search the separability for the other integrable cases 
it is convenient to use the spheroidal coordinates
$(\xi,\eta,\varphi)$, where $\xi = (r_1+r_2)/d$ and $\eta = (r_1-r_2)/d$.
In the {\it prolate} spheroidal coordinates
$r_1 = [\rho^2+(z+d)^2]^{1/2}$, $r_2 = r$. 
The parameter $d\in(0,\infty)$ is the distance between two foci of the
coordinate system (with the origin at one of them). 
In the limit $d\to 0$ the motion is separated when
 $\omega_z/\omega_\rho=1$ (Fig.~\ref{fig2}b). In this limit
$\xi\to\infty$ such that
$r=d\xi/2$ is finite, $\eta = \cos\vartheta$, and we obtain the spherical
coordinate system. 

Let us turn to the case $\omega_z/\omega_\rho = 2$ which
occurs at the value of the magnetic field
$\omega_L^{\prime \prime} = (\omega_z^2/4 - \omega_0^2)^{1/2}$. 
In the prolate  spheroidal coordinates the motion is 
separated in the limit $d\to\infty$ (Fig.~\ref{fig2}d).
In fact, at $d\to\infty$: $\xi\to 1$, $\eta\to 1$ such that
$\xi_1 = d(\xi-1)$, $\xi_2 = d(1-\eta)$ are finite, -- we obtain the
parabolic coordinate system $(\xi_1,\xi_2,\varphi)$ where
$\xi_{1,2} = r\pm z$. In these coordinates 
the Hamiltonian (\ref{relham}) has the form
\bea
h \!&=&\! \frac{1}{\xi_1+\xi_2} \left[\,2(\xi_1 p_{\xi_1}^2 + 
\xi_2 p_{\xi_2}^2) +
\frac{m^2}{2}\left(\frac{1}{\xi_1} + \frac{1}{\xi_2} \right)
\right.
\nonumber \\[1ex]
\!&+&\! \left. 
\frac{{\tilde\omega_z}^2}{8}\,(\xi_1^3 + \xi_2^3) + \lambda \,\right] -
{\tilde\omega_L^{\prime \prime}}m
\label{relpar}
\eea
and the equation $(\xi_1+\xi_2)(h - \epsilon) = 0$ is separated into two
decoupled equations for $\xi_1$ and $\xi_2$ variables.
Simple manipulations define the separation constant 
\beq
c = a_z - {\tilde\omega_{\!\rho}}^2\rho^2 z
\label{ni}
\eeq
which is a desired third integral of motion.
Here $a_z$ is the z-component of the Runge-Lenz vector
${\mathbf a} = {\mathbf p}\times{\mathbf l} +
\lambda{\mathbf r}/(2r)$, which is a constant of motion for
the pure Coulomb system 
(i.e. when $\omega_{\rho} = \omega_z = 0$) \cite{LL}.
The quantum mechanical counterpart of the
integral of motion, Eq.~(\ref{ni}), does not commute
with the parity operator and we should expect the degeneracy 
of quantum levels.

Due to the separability of the motion in the 
parabolic coordinate system,
the eigenfunctions of the corresponding Schr\"odinger
equation can be expressed in the form 
$\psi({\mathbf r}) = f_1(\xi_1)\,f_2(\xi_2)\, e^{im\varphi}$,
where the functions $f_j$ are solutions of the equations
\bea
\frac{d}{d\xi_j}\bigg(\xi_j\frac{d f_j}{d\xi_j}\bigg)
\!\!&-&\!\! \frac{1}{4} \bigg[ \frac{m^2}{\xi_j}+
\frac{\tilde\omega_z^2}{4}\xi_j^3 - 
2(\epsilon+\tilde\omega_L^{\prime\prime} m)\xi_j
\nonumber \\
\!\!&+&\!\!
\lambda-(-1)^j 2c\, \bigg]\,f_j = 0 ,\quad j=1,2.
\label{fgeqs}
\eea
Eqs.~(\ref{fgeqs}) can be solved numerically and the
eigenenergies and eigenvalues of $c$ are determined iteratively by varying
simultaneously $\epsilon$ and $c$ until the functions $f_j$
fulfill the boundary conditions:
$f_j \sim \xi_j^{|m|/2}$, $\xi_j df_j/d\xi_j = 
|m|f_j/2$ for $\xi_j \to 0$ and $f_j \to 0$ for $\xi_j
\to \infty$.
Let $n_1$ and $n_2$ be the nodal quantum numbers of the functions 
$f_1$ and $f_2$, respectively.
Note that Eqs.~(\ref{fgeqs}) are coupled by the constants of motion and, 
therefore, both functions depend on all three quantum numbers $(n_1,n_2,m)$. 
However, the simple product of these functions
have no a definite parity. Since
${\mathbf r} \to -{\mathbf r} \Leftrightarrow \{\xi_1\to\xi_2, \xi_2\to\xi_1,
\varphi\to\varphi + \pi\}$, the even/odd eigenfunctions are
constructed as
\bea
\psi^{(\pm)}_{N,k,m}({\mathbf r}) \!&=&\! 
\frac{e^{im\varphi}}{\sqrt{2}}\,
[f_{n_1}^{(n_2,m)}(\xi_1)\,f_{n_2}^{(n_1,m)}(\xi_2)
\nonumber \\[1ex]
\!&\pm&\! (-1)^m f_{n_2}^{(n_1,m)}(\xi_1)\,f_{n_1}^{(n_2,m)}(\xi_2)],
\label{psipar}
\eea
where $N = n_1+n_2$ and $k = \vert n_1-n_2\vert$.
These states 
are the eigenfunctions of $h$, $l_z$, $|c|$ and the parity operator.
For $|c| > 0$  the eigenstates (\ref{psipar}) 
appear in doublets of different parity and, therefore, of a different 
total spin.
For  $c = 0$ in Eqs.~(\ref{fgeqs}) $f_1 = f_2$ and, obviously, 
only the states with parity $\pi = (-1)^m$ exist.

For the magnetic field 
$\omega_L^{\prime\prime\prime} 
\equiv (4\omega_z^2 - \omega_0^2)^{1/2}$ we obtain 
the ratio $\omega_z/\omega_\rho = 1/2$.
The Hamiltonian (\ref{relham}) expressed
in the {\it oblate} spheroidal coordinates 
($r_1 = [z^2+(\rho+d)^2]^{1/2}$, $r_2 = r$)
is separated for $m = 0$ (at $d\to\infty$). 
For $m \neq 0$ the term
$m^2/\rho^2$ and, consequently, the Hamiltonian (\ref{relham}) 
is not separated in these coordinates. Note, for $m = 0$ the cases
$\omega_z/\omega_\rho = 1/2$ and $2$ are equivalent
if we exchange the $\rho$ and $z$ coordinates and,
hence, the additional integral of motion is 
$\vert a_\rho - {\tilde\omega_{\!z}}^2 z^2\rho \vert$.
For $m\neq 0$ we use 
the procedure described in Ref.~\cite{blum} and obtain the
following integral of motion 
\beq
C = [(a_\rho - {\tilde\omega_z}^2 z^2\rho)^2 + a_\varphi^2 +
4m^2\tilde\omega_z^2 r^2]^{1/2},
\label{secint}
\eeq
where $a_\rho$ and $a_\varphi$ are the $\rho$ and $\varphi$
components of the Runge-Lenz vector, respectively. Due to
existence of three independent integrals of motion, $h$, $m$ and $c$,
which are in involution, the dynamics for $m \neq 0$, although
non-separable, is integrable. 
The further analysis for $m=0$ is similar to the
previous one and we omit it here. 

Let us denote by $h^*$ the Hamiltonian (\ref{relham}) for a specific value of
the magnetic field when the system becomes separable, i.e. 
for $\omega^*_{\!L} =  \omega^\prime_L$, $\omega^{\prime\prime}_L$ or
$\omega^{\prime\prime\prime}_L$ (for $m=0$). 
Then for an arbitrary value of $\omega_L$ we can write
\beq
h = h^* + \hbox{$\frac{1}{2}$} \Delta{\tilde\omega_{\! L}}^2 \rho^2 -
\Delta{\tilde\omega_{\! L}}m,
\label{relham1}
\eeq
where $\Delta{\tilde\omega_{\! L}}^2 = {\tilde\omega_{\! L}}^2 - 
{\tilde\omega}_L^{*2}$, $\Delta\tilde\omega_L = {\tilde\omega_L} -
{\tilde\omega_{\! L}^*}$ and
the term $\frac{1}{2} \Delta{\tilde\omega_{\! L}}^2\rho^2$ is the
only non-diagonal part of $h$ in the eigenbasis of $h^*$. 
The eigenenergies of the Hamiltonian (\ref{relham}) (see Fig.~\ref{fig3}) 
have been calculated with the use
of the basis (\ref{psipar})
and the spherical basis 
in the intervals $0 \le \tilde\omega_L \le 1.5$ and the
$1.5 \le \tilde\omega_L \le 5$, respectively.
The radial parts of the spherical eigenfunctions 
and $f_{n_1,n_2,m}$, as well as the
corresponding factors in the matrix elements $\langle \psi_i \vert\, \rho^2
\vert \psi_j \rangle$, are evaluated numerically.
The complete spectrum of the two-electron QD (Fig.~\ref{fig3}a) shows the
accumulation of levels with different quantum numbers 
into well pronounced bands at strong magnetic field. 
There is no obvious manifestation of the 
symmetries discussed above. In fact, the effects of symmetries are shown up 
for separated $m$-manifolds only (Fig.~\ref{fig3}b).

For non-interacting electrons ($\lambda = 0$), the energy
levels of the QD are Fock-Darwin levels \cite{Fock}
\beq
\epsilon = \tilde\omega_\rho (2 n_\rho + \vert m\vert + 1) +
\tilde\omega_z (n_z + \hbox{$\frac{1}{2}$}) - \tilde\omega_L m.
\label{fock}
\eeq
For rational ratios of $\omega_z/\omega_\rho$ the energy 
levels Eq.~(\ref{fock}) are degenerate. 
It is simply the spectrum of the RHO in 
the external field $\tilde\omega_L m$.
For instance, at $\omega_z/\omega_\rho = 2$ we have
$\epsilon = \tilde\omega_z (N + \vert m\vert/2 + 1) -
\tilde\omega_L^{\prime\prime} m$. The quantum number  
$N = n_\rho + n_z = n_1 + n_2 = 0, 1, 2, ...$ and 
each $m$-manifold consists of the shells characterized
by this quantum number. 
Since the eigenenergies of  the term $H_{\mathrm CM}$
with the corresponding quantum numbers
are determined by Eq.~(\ref{fock}) as well, 
the shells in the total spectrum of the QD  
are not affected.

The Coulomb interaction destroys the general symmetry
of the 3D HO. However, the magnetic field can recover 
symmetries which are common for 
the RHO and Coulomb systems.
At a relatively low value  
of the magnetic field $\omega_L^{\prime \prime}$  
(for our parameters $B\approx2.4$\,T) we reveal
the first manifestation of the hidden symmetries. 
This symmetry is determined by the integral 
of motion, Eq.~(\ref{ni}).  
It results in the appearance of shells at
each $m$-manifold (Fig.~\ref{fig3}b). 
There are exact crossings and repulsions 
between  levels of different 
and of the same parity, respectively, in each shell. 
The near-degeneracy of the quantum spectrum 
is reminiscent of a striking degeneracy observed for 
the RHO or pure Coulomb systems. 
At higher values of the magnetic field $\omega^\prime_L$  
($B \approx 7.5$\,T), the dynamical
spherical symmetry appears, since ${\bf l}^2$ 
becomes an additional integral of motion.
This symmetry manifests itself as the attraction between levels 
with different orbital quantum numbers and the same 
parity (Fig.~\ref{fig3}b). In contrast to spectra of pure Coulomb 
systems or of the RHO, there are no crossings
between eigenstates of the subset characterized by a given 
quantum number $m$, since the accidental degeneracy is
removed. 
Although the symmetry is recovered at 
very strong magnetic field $\omega_L^{\prime \prime\prime}$
($B\approx 15.9$\,T)
due to the appearance of the integral of motion Eq.~(\ref{secint}),
the dynamics is non-separable for $m\neq0$. Note that shells are similar 
to the spherical case.

The symmetries may be detected by studying the 
conductance of two-electron QDs at low temperatures.
In particular,
at $\omega_L^{\prime \prime}$ in the excited states
there is the onset of a singlet-triplet degeneracy
related to crossings of the eigenstates (\ref{psipar}) 
with $\vert c\vert > 0$ (see Fig.~\ref{fig3}b).
The total spin $S$ alternates between 1 and 0 
and the addition of a second electron with 
a spin-up or spin-down orientation 
to the QD will cost the same energy.
At zero magnetic field $B=0$ two electrons occupy the same state
with $S=0$. At $\omega_L^{\prime \prime}$ 
it becomes favorable for one electron to flip its spin.
The electron reconfigures the charge
and polarizes the two-electron QD leading to
a Kondo type effect \cite{Kon}, i.e., to the increase of the
conductance at low temperatures.
The enhancement of the conductance in QDs 
due to the singlet-triplet degeneracy induced by 
the magnetic field have been discussed in \cite{Meir,Pus,Glaz} 
(for review see \cite{revP}).
According to Ref.\cite{revP}, when the system is tuned to the
degeneracy point the differential conductance, 
$dI/dV$, would exhibit a peak at zero bias. 
The increase of the strength of the magnetic field 
removes the degeneracy
and the peak will split onto two peaks reflecting the
single-particle spacing between singlet-triplet states. 
It should be noted, however, that this prediction  is obtained
 in Ref.\cite{revP} within a schematic model 
 where single-particle levels and the
magnetic field are adjustable parameters. In addition, 
the electron--electron interaction  
is assumed to be weak.
According to our analysis the onset of 
the singlet-triplet degeneracy holds at any
strength of the electron-electron interaction in QDs. 
We suggest the mechanism, related to 
the hidden symmetries,
responsible for the occurrence of
this degeneracy in two-electron 3D QDs
at a certain value of the magnetic field.

Summarizing, we have shown that 
the axially symmetric 3D quantum dot with two electrons 
exhibits hidden symmetries 
at certain values of the magnetic field. 
In particular, due to these symmetries 
the onset of a singlet-triplet degeneracy 
in excited states is found
when the magnetic field value is 
$\omega_L^{\prime \prime}$.
Finally, we are thankful to Jan--M.~Rost for valuable discussions and
constructive suggestions.

\newpage
\centerline{\bf Figure Captions}

\vskip 2cm

\begin{figure}
\caption{
Poincar\'e surfaces of sections $z = 0$, $p_z > 0$ 
of the relative motion ($\lambda=3$, $\epsilon = 10$, $m = 0$) with: 
(a) $\omega_z/\omega_\rho = 5/2$,
(b) $\omega_z/\omega_\rho = 2$ and
(c) $\omega_z/\omega_\rho = 3/2$.
The section (b) indicates that for the ratio
$\omega_z/\omega_\rho = 2$ the system is integrable.}
\label{fig1}
\end{figure}

\begin{figure}
\caption{
Typical trajectories ($\epsilon = 10$, $m = 1$) of the
relative motion at $\lambda=3$
for $\omega_z/\omega_\rho = 1$ (a,b) and $\omega_z/\omega_\rho = 2$ (c,d)
are shown in cylindrical and prolate spheroidal coordinates, respectively.
}
\label{fig2}
\end{figure}

\begin{figure}
\caption{The lowest eigenenergies of  
Hamiltonian (\ref{relham})
(in units $\hbar\omega_0$) at $\omega_z/\omega_0 = 2.5$
and $\lambda=3$ as functions of the ratio $\omega_L/\omega_0$ for: 
(a) all $m$;
(b) $m = 0$. 
The upper energy limit is chosen high enough to amplify
the shell structure of the spectrum.
The integrable cases: $\omega_z/\omega_\rho = 2$,
1 and 1/2 are indicated by vertical dotted lines
($\omega_L/\omega_0 = 0.75$, $\sqrt{21}/2$ and $\sqrt{24}$, respectively).
}
\label{fig3}
\end{figure}

\end{document}